\documentclass[12pt,aps,prl,preprint,tightenlines]{revtex4} 
\usepackage{amsfonts,amssymb,amsmath,amscd}

\begin{document}

\title{A  logical loophole in the derivation of the CHSH Bell-type inequality} 

\author{Gerrit Coddens\\
Laboratoire des Solides Irradi\'es,\\
CEA-DSM-IRAMIS, CNRS~UMR~7642,~Ecole Polytechnique,\\
28, Route de Saclay, F-91128-Palaiseau CEDEX, France}

\date{Received: date / Revised version: date}
\begin{abstract}
We point out a loophole in the derivation of  the Bell inequalities in the form proposed by 
Clauser, Horne, Shimony and Holt (CHSH). In this derivation it is assumed that
statistical independence is a necessary consequence of locality, but this is not a logical necessity.
\end{abstract} 


\maketitle



One of the most salient features that emerges from the history
 of the study of the foundations
of quantum mechanics (QM) is how the 
formulation of the EPR paradox  \cite{EPR}
led to the derivation of the Bell-type inequalities  \cite{Bell}
 and the brilliant experiments
of Aspect {\em et al.}  \cite{Aspect,Aspect1}.
The comparison of the experimental results with the inequalities
shows that they violate the inequalities, which leads to the conclusion that QM cannot be a hidden-variables
theory. In the present article we question the generality of the validity of
the Bell inequalities used in the context of these experiments by showing that they are built on an assumption that is not logically cogent.

The essentials of the subject matter have been expounded in great clarity by Shimony
\cite{Shimony}.
In the experiments of Aspect {\em et al.} one considers a composite system
of two particles. The first particle impinges upon an apparatus $\alpha$ with an adjustable parameter A (a device that
can take  two orientations
$A_{1}$ and $A_{2}$). 
Two complementary outcomes, labeled $\oplus$ (the particle is transmitted by the device and registered in a detector behind the device)
and $\ominus$  (the opposite is true) are possible for each device setting. 
 In a completely analogous way, 
the second particle travels to another apparatus $\beta$. The adjustable
 parameter is now called B (also a device that
can take  two orientations
$B_{1}$ and $B_{2}$), and again two complementary results, $\oplus$ and $\ominus$ are possible for each device setting.
In the following we will also use the terminology detector to refer to an apparatus.

The probabilities of the $\oplus$ readings are noted by
$p(A)$ and $p(B)$. The probabilities for the $\ominus$ results are then
$1-p(A)$ and $1-p(B)$ respectively. 
These probabilities are compared  with a very general Bell-type inequality.
Shimony \cite{Shimony} explains how one can derive such a very general inequality for
 probabilities that the particles have properties $A_{1}$, $A_{2}$, $B_{1}$ and $B_{2}$.
 The derivation of the
inequality starts from a generally valid algebraic inequality
for any four numbers $(r_{1},r_{2},s_{1},s_{2}) \in [0,1]^{4}$, that can be easily
checked on a Venn diagram in set theory:

\begin{equation} \label{1}
-1 \le r_{2}s_{2} + r_{2}s_{1} + r_{1}s_{2} -r_{1}s_{1} -r_{2} -s_{2} \le 0
\end{equation}

\noindent From this one derives for  probabilities $(p(a_{1}),p(a_{2}),p(b_{1}),p(b_{2}))
 \in [0,1]^{4}$:

\begin{equation}\label{2}
-1 \le p_{\lambda}(a_{2}\cap b_{2}) + p_{\lambda}(a_{2} \cap b_{1}) +
 p_{\lambda}(a_{1} \cap b_{2}) - p_{\lambda}(a_{1} \cap b_{1})
 - p_{\lambda}(a_{2}) - p_{\lambda}(b_{2}) \le 0
\end{equation}

\noindent  where $p_{\lambda}(a \cap b)$ denotes the probability for
 the joint outcome of $a$ and $b$. The index $\lambda$ is used to indicate
that the probabilities might depend on a number of parameters, over which one still has to integrate using a distribution
 $\rho(\lambda)\,d\lambda$,
which yields the final result:

\begin{equation}\label{3}
-1 \le p(A_{2} \cap B_{2}) + p(A_{2} \cap B_{1}) + p(A_{1} \cap B_{2}) 
- p(A_{1} \cap B_{1})
 - p(A_{2}) - p(B_{2}) \le 0
\end{equation}

\noindent  where $p(A \cap B)$ denotes now the probability for
 the joint outcome of $A$ and $B$. This is the CHSH inequality.
It is this inequality that was reported to be violated by QM.
 
The derivation of the inequality  is based on an assumption of statistical
independence.
In the derivation of equation (2) from equation (1) statistical independence  is
 translated by $p_{\lambda}(a\cap b) =
p_{\lambda}(a)\,p_{\lambda}(b)$ and is used in order
 to derive equation (2)
from equation (1). In fact, the real expression should read
$p_{\lambda}(a\cap b) = p_{\lambda}(a)\,p_{\lambda}(b\, |\, a)$, where
the ``conditional'' probability $p_{\lambda}(b \,|\, a)$ for the occurrence of $b$ provided
$a$ has occurred (where for mnemonics $|$ could be read as
 ``provided that'') is identified with 
$p_{\lambda}(b)$, based on the independence conditions.

 By making the distance between the detectors large enough one  makes sure that the experimental results are Einstein-independent. 
 The outcome of an experiment in detector $\alpha$ cannot influence the outcome of an experiment
 in detector $\beta$. The information obtained at detector $\alpha$ would have to travel faster than light to be available
 at detector $\beta$ when the measurement is carried out at detector $\beta$. This expresses the principle of locality,
 which is a basic ingredient of realism. It implies that the experiments are carried out in a double-blind fashion.
 We use the terminology ``Einstein independence'' for locality, in order to be able to confront it with
 statistical independence.
 
CHSH have used the assumption of statistical  independence to express Einstein independence: Both types of independence
 express that what goes on in detector $\alpha$ must be completely independent of what goes on in detector $\beta$.
We can summarize the  assumption of CHSH as the implication ${\cal{I}}$:  ``Einstein independence $\Rightarrow$ statistical independence''.
  It is this statistical independence that is used in the derivation of the inequalities and that is needed to justify the step 
  from Eq. \ref{1} to Eq. \ref{2}. 
  It is the Einstein independence that is warranted by the design of the experiments.
  There is no further discussion of the implication in the literature, as though it would be self-evident.
 But we will now show that from a logical viewpoint the implication ${\cal{I}}$ is not necessarily true.
 Let us show this  by a counter-example of a rather abstract {\em Gedankenexperiment}. 
 The point of this {\em Gedankenexperiment} is not to describe a true possible experiment, but to provide a logical counter-example
 permitting to point out that  the implication ${\cal{I}}$ is not generally valid.
 
 Imagine that we have pairs of  two identical reference frames $X_{1}$, $X_{2}$. We can consider these frames as attached to
 two identical particles in the experiment. We will just use these frames to describe the properties of the particles. These properties will be rotational properties.
 E.g. we could imagine that the polarization or the spin of a particle is rotated in the interaction with a device. The frames attached to the particles
 can then be used to describe these changes in orientation by co-rotating with the spin or the polarization.
 Or we could imagine that the interaction with the device depends on the orientation of the spin or the polarization of the particle with
 respect to the device. We can then use the frame $X$ to describe that orientation.
 The two frames are identical in that they have the same orientations. This could e.g. translate the fact that the spins of the two particles are identical.
 For the rest, we assume the pair can have
all possible different orientations as a pair. We can specify these orientations e.g.  through the triplet of Euler 
angles ${\boldsymbol{E}}_{X} = (\alpha_{X},\beta_{X},\gamma_{X})$ with respect to a fixed reference frame $F$  in the laboratory.
In fact, each orientation of a frame $X$ corresponds in a 1-1 way to a rotation $R_{X}$ with respect to this fixed frame $F$, {\em viz.} the rotation that is needed to make the orientation
of the fixed frame $F$ coincide with that of the frame $X$: $X=R_{X}(F)$. When we say that the frames $X_{1}$ and $X_{2}$ can have all orientations as a pair, this means thus that
$R_{X}$ can be any element of the rotation group SO(3) and has a uniform distribution over SO(3).

The two particles travel with their frames without changing their orientation in opposite directions. 
Particle $1$  flies with its frame $X_{1}$  to a device $\alpha$. 
To describe the orientation of the device $\alpha$ in space we use a frame
$A$ attached to it. We could need this frame e.g. to describe the orientation of a polarizer in space.
The orientation of frame $A$ is described by the Euler angles ${\boldsymbol{E}}_{A} = 
(\alpha_{A},\beta_{A},\gamma_{A})$ with respect to $F$. We note the rotation that is in 1-1-correspondence with
the frame  $A$ as $R_{A}$, i.e. $A= R_{A}(F)$.
Particle $2$ flies with  its frame $X_{2}$  to a device $\beta$, that  also has a reference frame $B$ attached to it to describe 
its orientation.   The device $\beta$ is identical in construction to device $\alpha$, and the frame $B$ is attached to $\beta$ in an exactly
identical way as $A$ to $\alpha$. Device $\beta$
may be just orientated differently in space than device $\alpha$.
The orientation of frame $B$ is described by the Euler angles ${\boldsymbol{E}}_{B} = (\alpha_{B},\beta_{B},\gamma_{B})$ 
with respect to $F$.
We note the rotation that is in 1-1-correspondence with
the frame  $B$ as $R_{B}$, i.e. $B=R_{B}(F)$.

We assume now that  what happens to the first particle with its attached frame $X_{1}$ at the device $\alpha$ 
is uniquely defined by the Euler angles ${\boldsymbol{E}}_{XA} = (\alpha_{XA},\beta_{XA},\gamma_{XA})$
 of the rotation
$R_{XA}$ that is needed to make the orientation of $X_{1}$ coincide with the orientation
of $A$.  That means thus that only the relative orientation of the particle with respect to the device
plays a r\^ole in the probabilities for what happens to the particle. This rotation is given by: $R_{XA}= R_{A} ~^{\circ}R_{X}^{-1}$ (Here the symbol $^{\circ}$ stands for the composition of functions). 
We can write the probability for what happens thus as $p_{\lambda}(a) = p({\boldsymbol{E}}_{XA})$. 
All the parameters needed to define the probabilities are thus present in ${\boldsymbol{E}}_{XA}$, which is defined by
${\boldsymbol{E}}_{X}$ and ${\boldsymbol{E}}_{A}$,
such that the probabilities are locally defined.
We may note that it is not compulsory that all three angles ${\boldsymbol{E}}_{XA}$ are necessary to define
the probability, but it is logically conceivable. We will assume that all three angles intervene.

We also assume that what happens to $X_{2}$ at the device $\beta$ is uniquely defined by the Euler angles ${\boldsymbol{E}}_{XB} = (\alpha_{XB},\beta_{XB},\gamma_{XB})$ 
of the rotation
$R_{XB}$ that is needed to make the orientation of $X_{2}$ coincide with the orientation
of $B$. This rotation is given by: $R_{XB}= R_{B} ~^{\circ}R_{X}^{-1}$. 
Again, this expresses that only the relative orientation of the particle with respect to the device
plays a r\^ole in the probabilities for what happens to the particle. This is logical as the two devices are identical such 
that they should work in the same way on identical particles..
We can write the probability for what happens  thus as $p(b) = p({\boldsymbol{E}}_{XB})$. 
All the parameters needed to define the probabilities are thus present in ${\boldsymbol{E}}_{XB}$, which is defined by
${\boldsymbol{E}}_{X}$ and ${\boldsymbol{E}}_{B}$,
such that the probabilities are locally defined.
Here again it is not compulsory that all three angles ${\boldsymbol{E}}_{XB}$ are necessary to define
the probability, but it is logically conceivable, and we will assume that all three angles intervene.

These conditions imposed on the probabilities imply that all physical processes we consider are local. 
There is nothing we need to know about what is going on at apparatus  $\beta$
to decide what has to be done in apparatus $\alpha$, and {\em vice versa}. The processes are thus Einstein independent.
The reader will notice that this {\em Gedankenexperiment} is inspired  by the set-up of the experiments of Aspect {\em et al.}
However, there is no need to reproduce any real experiment at all. We are just investigating a logical possibility.

Now CHSH have assumed that the implication ${\cal{I}}$ is true to make the step
from Eq. \ref{1} to Eq. \ref{2}. Without this assumption the derivation breaks down.
We will show now that ${\cal{I}}$ is not a logical necessity.

Let us ask if the mathematical expression for the quantity 
$p_{\lambda}(b|a) = p({\boldsymbol{E}}_{XB}|{\boldsymbol{E}}_{XA})$ 
is independent from the mathematical expression
for the quantity $p_{\lambda}(a) = p({\boldsymbol{E}}_{XA})$. The following argument shows that this is not the case. 
In fact, the three  Euler angles ${\boldsymbol{E}}_{XB}$ are not independent from the Euler angles ${\boldsymbol{E}}_{XA}$.
There is a geometrical correlation between the two because there
 is a unique well-defined rotation $R_{AB} =  R_{B} ~^{\circ}R_{A}^{-1}$
that permits to rotate $A$ to $B$. In other words, there is a relationship between the Euler angles 
${\boldsymbol{E}}_{XA}$
and ${\boldsymbol{E}}_{XB}$, because $R_{XB} = R_{B} ~^{\circ}R_{X}^{-1} = R_{B} ~^{\circ}R_{A}^{-1} ~^{\circ} R_{A} ~^{\circ}R_{X}^{-1} = R_{AB} ~^{\circ} R_{XA}$. 
In fact,  using the identity $R_{XB} = R_{AB} ~^{\circ} R_{XA}$, the Euler angles
${\boldsymbol{E}}_{XB}$ can be completely
expressed in terms of the Euler angles ${\boldsymbol{E}}_{XA}$ and ${\boldsymbol{E}}_{AB}$, and to make the
calculations correctly we must do this: 
${\boldsymbol{E}}_{XB} = f({\boldsymbol{E}}_{XA},{\boldsymbol{E}}_{AB})$, where $f$ is the function that defines the relationship.
The function that expresses ${\boldsymbol{E}}_{XB}$
in terms of ${\boldsymbol{E}}_{XA}$, using the auxiliary parameters ${\boldsymbol{E}}_{AB}$  is even bijective.
The only exact way to calculate
 $p(A\cap B)$ is thus:
 
\begin{equation}
p(A\cap B) = \int_{R_{X} \in SO(3)}\, p[\,{\boldsymbol{E}}_{XA}\,]\, p[\,f({\boldsymbol{E}}_{XA},{\boldsymbol{E}}_{AB})\,]\,d{\boldsymbol{E}}_{XA}.
\end{equation}

\noindent Here $d{\boldsymbol{E}}_{XA}$ is a volume element to integrate over the rotation group. 
(It corresponds to $\rho(\lambda)d\lambda$ and is the volume element 
of the so-called Haar integral expressed in the Euler angles. Its value is ${1\over{8\pi^{2}}}\, \sin\beta_{XA}\, d\alpha_{XA} \,d\beta_{XA} \,d \gamma_{XA}$, with $(\alpha_{XA},\beta_{XA},\gamma_{XA}) \in [0,2\pi] \times [0,\pi] \times [0,2\pi]$, but this is immaterial for the logic of the argument).
The variable $R_{X}$ plays here the r\^ole of $\lambda$. We can see that the second term under the integral, which corresponds to $p_{\lambda}(b|a) = p({\boldsymbol{E}}_{XB}| {\boldsymbol{E}}_{XA})$
 in the derivation of the inequalities,
  is not mathematically independent 
of ${\boldsymbol{E}}_{XA}$. It contains ${\boldsymbol{E}}_{XA}$ and must so in order to express the 
existence of a well-defined relative orientation of frame $B$ with respect to frame $A$. It really
must because we integrate over ${\boldsymbol{E}}_{XA}$ such that ${\boldsymbol{E}}_{XB}$ must be expressed
in terms of ${\boldsymbol{E}}_{XA}$, and this expression will in general not be a constant but a function 
of ${\boldsymbol{E}}_{XA}$ (as we have
assumed that all Euler angles occur in the expressions for the probabilities $p({\boldsymbol{E}}_{XB})$).
Therefore $p({\boldsymbol{E}}_{XB}| {\boldsymbol{E}}_{XA}) = p({\boldsymbol{E}}_{XB})$ is not true.
In our set-up, the statistical-independence criterion is thus not satisfied
as assumed in the derivation of the Bell inequalities, despite the fact that Einstein independence is satisfied. 

Statistical independence is thus not a necessary consequence of locality (or Einstein independence).
In our {\em Gedankenexperiment} we have thus $\neg$(``Einstein independence $\Rightarrow$ statistical independence''), such that
the assumption that CHSH  made is not generally valid.
  It is pointless to quibble if our {\em Gedankenexperiment} could
describe the experiments of Aspect {\em et al.} or otherwise, because what we want to point out is a {\em logical error}. 
The Bell inequalities are built only on logic, without any input about physical issues. 
Exactly because they did
not address physical issues the Bell inequalities have been claimed to be completely general.
It is on the same battle field of pure logic without any dependence on physical issues, 
that we want to point out an error in this purely logical framework of the CHSH inequality, by giving a counter example showing that the  assumption
used to derive it is not a logical necessity. 
Our {\em Gedankenexperiment} may not correspond to a real physical experiment but it could be used make a
Monte Carlo simulation that could serve as an example for a case wherein the CHSH  assumption is not true,
 such that the assumption is not a logical necessity. As a consequence of the results of the experiments, the principle of locality has become
 under fire. But we can see that this is only due to a wrong transcription of the principle into the formalism using ${\cal{I}}$.

\end{document}